\documentclass{article}
\usepackage{spconf,amsmath,graphicx}
\usepackage{hyperref}
\usepackage{color}
\usepackage{multirow}

\usepackage{todonotes} 

\def\emo{IEmoNet}  

\makeatletter
\newcommand{\printfnsymbol}[1]{%
  \textsuperscript{\@fnsymbol{#1}}%
}
\makeatother
\title{Bimodal Speech Emotion Recognition Using Pre-trained Language Models}
\name{Verena Heusser\sthanks{equal contribution}, Niklas Freymuth\printfnsymbol{1}, Stefan Constantin, Alex Waibel}
\address{Karlsruhe Institute of Technology\\
Institute for Anthropomatics and Robotics\\
\{verena.heusser, niklas.freymuth\}@student.kit edu, \{stefan.constantin, waibel\}@kit.edu
}
\begin{document}
%
\maketitle
\begin{abstract}
Speech emotion recognition is a challenging task and an important step towards more natural human-machine interaction.
We show that pre-trained language models can be fine-tuned for text emotion recognition, achieving an accuracy of $69.5\,\%$ on Task 4A of SemEval 2017, improving upon the previous state of the art by over $3\,\%$ absolute.
We combine these language models with speech emotion recognition, achieving results of $73.5\,\%$ accuracy when using provided transcriptions and speech data on a subset of four classes of the IEMOCAP dataset. The use of noise-induced transcriptions and speech data results in an accuracy of $71.4\,\%$.
For our experiments, we created \emo \hspace{0.01cm}, a modular and adaptable bimodal framework for speech emotion recognition based on pre-trained language models.
Lastly, we discuss the idea of using an emotional classifier as a reward for reinforcement learning as a step towards more successful and convenient human-machine interaction.
\end{abstract}

\begin{keywords}
Speech Emotion Recognition, Text Emotion Recognition, Bimodal Emotion Recognition, IEMOCAP, Self Attention, Pre-trained Language Models
\end{keywords}  

\section{Introduction}
\label{sec:intro}
Emotions are an important aspect of human behavior. They do not only influence the reaction to our environment \cite{DeMartino2006frames, Bouffard2002sex}, but also actively change our perception of it \cite{Phelps2006Emotion} and sometimes even contribute to how well we remember specific events \cite{Kensinger2007negative}.
As such, they influence both human-human and human-machine interaction. However, in human-machine interaction, emotions are often not at all or only scarcely considered. 
We propose to use automatically generated emotional feedback as a reward in Reinforcement Learning (RL). This could be used to improve human-machine interaction, for example when used with dialogue systems. In this paper, we will focus on the tasks of automatic Text Emotion Recognition (TER) and automatic Speech Emotion Recognition (SER).
\par

It has been shown that combining textual features and a SER model into a multimodal system improves the overall classification performance \cite{Aguilar2019multimodal, Cho2018multimodal, Tripathi2018multimodal, Yoon2018multimodal}.
Additionally, the tasks of Automatic Speech Recognition (ASR) and language modelling have seen great improvements over the last year.
This leads to a steady improvement in the quality of both transcripts and language models.
Since most language models have been shown to be able to solve a plethora of different tasks \cite{Radford2019language, Liu2019mtdnn}, the combination of ASR and TER can be used to automatically transcribe a given utterance and use the language model for an emotion classification.
To show the effectiveness of language models for TER, we fine-tune both BERT \cite{Devlin2019bert} and XLNet \cite{Yang2019xlnet}, two self-attentive language models, on Task 4A of SemEval 2017 \cite{Rosenthal2017semeval}.
\par 
Motivated by these two recent advancements, we present \emo \hspace{0.01cm} (Interactive Emotion Network) which is a modular and adaptable bimodal SER framework based on pre-trained language models. \emo consist of independently trainable ASR, SER and TER sub-modules. They can be combined by fine-tuning only a few layers, omitting time-consuming re-training of the whole model. This allows us to utilize both textual and paralinguistic features for emotion classification.
We test our model on the `Text+Speech' subtask of the Interactive Emotional dyadic MOtion CAPture (IEMOCAP) dataset \cite{Busso2008iemocap}. Using 10-fold cross validation, \emo \hspace{0.01cm} achieves results of $73.5\,\%$ when combining text and speech inputs. Using speech and (simulated) automatically transcribed texts instead of the provided transcriptions results in $71.4\,\%$ accuracy.
\par 
\section{Related Work}
\label{sec:relatedWork}
\emo \hspace{0.01cm} is an approach to bimodal SER, meaning that it uses different modalities to extract richer and more varied features. In general, bi- and multimodal models combine modalities such as video recordings \cite{Poria2016multimodal}, more specific facial features \cite{Busso2004er}, movements \cite{Noroozi2018survey} and even EEG signals \cite{sreeshakthy2016brain}. In our case text, transcribed speech and speech are used, closely following \cite{Tripathi2018multimodal, Yoon2018multimodal, Xu2019multi}. 
Previous research focused on the Time Delay Neural Network (TDNN) \cite{Waibel1989tdnn} and recurrent architectures like the Long Short-Term Memory (LSTM) \cite{Hochreiter1997lstm}.\par 
Recently, ELMo \cite{Peters2018deep}, an LSTM-based pre-trained language model, has been successfully used to generate textual features for multimodal emotion recognition \cite{Aguilar2019multimodal}. 
We use pre-trained language models based on Transformer encoder stacks \cite{Vaswani2017attention} instead. These Transformer encoder stacks rely solely on the self-attention mechanism \cite{Bahdanau2014neural} for contextualization. They have shown great potential for ASR, SER and (emotional) language modelling individually, as will be shown in the following sections.

\subsection{Speech Emotion Recognition}
SER has been studied for several decades \cite{Chandrasekar2014survey}, with early work consisting of the classification of a given utterance to one of usually four distinct emotions based on prosodic features such as pitch, speech and intensity \cite{dellaert1996recognizing, Polzin1998Emotions}.
While these relatively simple features are still used today, deep neural architectures have led researchers to shift towards either more features such as the IS09 emotional feature set \cite{Schuller2009IS}, or more or less unprocessed audio/signal data as input \cite{ElAyadi2011emotionSurvey, Schuller2018speech}. \par 
Training and evaluating SER models is challenging because most datasets only include a few hours of spoken data, making it difficult to train large neural networks without severe overfitting. 
Research on SER utilizes Mel Frequency Cepstral Coefficients (MFCCs), spectrograms and raw waveforms as inputs, while the architectures vary from gated recurrent units \cite{Cho2014gru} to a combination of LSTMs and Time Delay Neural Networks (TDNNs) \cite{Sarma2018emotionIF, Satt2017efficientER}.
Recently, the Transformer has also been used in SER, either directly \cite{Tarantino2019transformer} or as a means of richer representations via predictive coding \cite{Lian2018transformer}, achieving competitive results in both cases.

\subsection{Automatic Speech Recognition}
While ASR faces similar challenges as SER, research has shown that transfer between ASR and SER models is only possible in early layers of the respective models, meaning that both tasks require their respective models to learn vastly different features \cite{Fayek2016OnTC}.
\par
As of today, a few architectures have shown to be well suited for ASR, all of which make either direct or indirect use of neural networks. 
Looking at the LibriSpeech corpus \cite{Panayotov_2015} as an example, Hybrid Models \cite{Yang2018hybrid}, Convolutional networks \cite{Li2019jasper} and LSTMs \cite{Chan2016LAS, Irie2019LAS, Park2019specaugment} all achieve Word Error Rates (WER) between $2.5\,\%$ and $3\,\%$ on the 'test-clean' sub-task when combined with a language model.
Additionally, researchers found that models based on the Transformer are also competitive and very promising, especially in regards to more complicated tasks \cite{Dong2018speechTransformer, Pham2019verydeep}. \par 

\subsection{Text Emotion Recognition}
TER is similar to sentiment analysis. Given an emotional model, usually Ekman's Base Emotions \cite{Ekman1999basic}, the task is to find the dominant emotion in a text. 
Common approaches are either keyword based or end-to-end. Keyword based methods operate on strong emotional words \cite{Goncalves2013sa} or are based around meta expressions like hashtags \cite{Kouloumpis2011hashtag}, emoticons \cite{Hogenboom2013emoticons}, and emojis \cite{Felbo2017deepmoji}. End-to-end architectures are mostly based on LSTMs and the attention mechanism \cite{Baziotis2017semeval, Cliche2017semeval}, allowing the models to correctly model long-term dependencies.\par 
However, with the introduction of attention-based models like the Transformer \cite{Vaswani2017attention} model and later BERT \cite{Devlin2019bert}, general language models are currently achieving great results on most Natural Language Understanding tasks, including sentiment analysis. 
An example is the GLUE Benchmark \cite{Wang2018glue}, a collection of various Natural Language Understanding tasks.
Since GLUE's release in 2018, BERT \cite{Devlin2019bert} and similar models \cite{Yang2019xlnet, Liu2019roberta, Zhu2019freelb, Lan2019albert} gave way to over a dozen improvements over the state of the art and eventually led to surpassing human performance on this benchmark\footnote{See \url{https://gluebenchmark.com/leaderboard} (accessed 2019-10-14)}.\par
Additionally, some of these language models are provided with pre-trained models and weights, allowing for efficient use of resources via transfer learning. These models can be fine-tuned on a given task, which means that some or all of the existing weights are trained on the new task for relatively few iterations. This drastically reduces both training time and the necessary amount of training samples, making it ideal for the data-sparse field of Emotion Recognition.
Utilizing this, researchers have achieved good results in both combining language models with LSTM architectures \cite{Huang2019semeval} and fine-tuning them on sentiment analysis tasks \cite{Xu2019Bert}, showing that they are suitable for the task at hand.
\par 

\subsection{Emotion in Reinforcement Learning}
In the context of dialogue systems (and human-machine interaction in general) it seems rather important to integrate the emotional state of the speaker into the overall decision making process of an agent. As of now, emotions have been used in both task-oriented \cite{Gnjatovic2008dialogue} and non task-oriented dialogue \cite{Chiba2018dialogue}, in both cases increasing the performance of the given system. Additionally, there has been a lot of research regarding emotionally intelligent systems which implement and operate based on their own emotional state \cite{Moerland2018RL}.\par 
\section{Reinforcement Learning}
\label{sec:rl}
Using emotion recognition for reinforcement learning in dialogue systems has the potential to greatly improve both the systems usefulness and the users overall experience. For example, a call center could track a caller's anger \cite{Burkhardt2017anger} and annoyance \cite{Irastorza2019tracking} levels to see how satisfied they are with their current call. In the case of a unsatisfactory calling experience, the emotional reaction can then be interpreted as a negative reward and used to adjust the dialogue system, causing it to slightly change its behaviour and react more appropriately in future calls.\par 
However, there is an important distinction between emotion and mood \cite{Berna2010induction, Sereno2015emotion}, the main difference being the timescale of both concepts \cite{Lane2005emotionMood}.
While emotion itself is short-term, its effects on a person's mood might persist for a relatively long time. In the context of dialogue systems, this means that a particularly good dialogue turn might be enough to cheer a person up and make them sound happier for the rest of the interaction. 
This again generates more `happy' features for the next turns, causing the SER system to classify what would normally be `neutral' as `happy'.
To counteract this effect of a `lingering' mood, we propose to use an approximated derivative of the emotional state instead of the state itself. For an emotion $e$ and a dialogue history $h_t$ at dialogue turn $t$, the output of the emotion recognition system can be seen as the posterior probability $P(e|h)$. The delta $\Delta P(e|h_t)=P(e|h_t)-P(e|h_{t-1})$ of each individual probability can then be seen as the resulting change in its corresponding emotion based on the last dialogue turn.\par 
Additionally, certain messages are expected to have certain emotional responses. Being the proverbial bearer of bad news, for example, a dialogue system should not perceive an anticipated negative emotion as a negative reward. Mathematically, this can be seen as the difference between the probabilities of the estimated emotional response $P^{est}(e|h_t)$ and the actual response $P(e|h_t)$ of the dialogue turn in question. Considering mood, the unexpected change becomes $\Delta P(e|h_t) - \Delta P^{est}(e|h_t).$\par 
Since a reward for reinforcement learning is usually bounded by a fixed interval, a mapping $f:E\rightarrow \lbrack p, q\rbrack$ between the set of emotions $E$ and their reward is necessary. In our case, a negative emotion like `anger' would be mapped to a negative value, the simplest case being $-1$. Taking this into account, the actual reward $r_t$ used for reinforcement learning would be
\begin{equation}
    r_t = \sum_{e\in E} f(e)\cdot (\Delta P(e|h_t) - \Delta P^{est}(e|h_t))\text{.}
\end{equation}
Given that it is easier to give an estimated reward rather than a set of changes in emotions, this can be simplified to
\begin{equation}
    r_t = \sum_{e\in E} f(e)\cdot \Delta P(e|h_t) -r^{est}_t\text{,}
\end{equation}
where $r^{est}_t$ is the estimated reward for dialogue turn $t$ and could be inferred from e.g. another neural network via regression.

\section{MODEL}
\label{sec:model}
\begin{figure}[htb]

\begin{minipage}[b]{1.0\linewidth}
  \centering
  \centerline{\includegraphics[width=8.5cm]{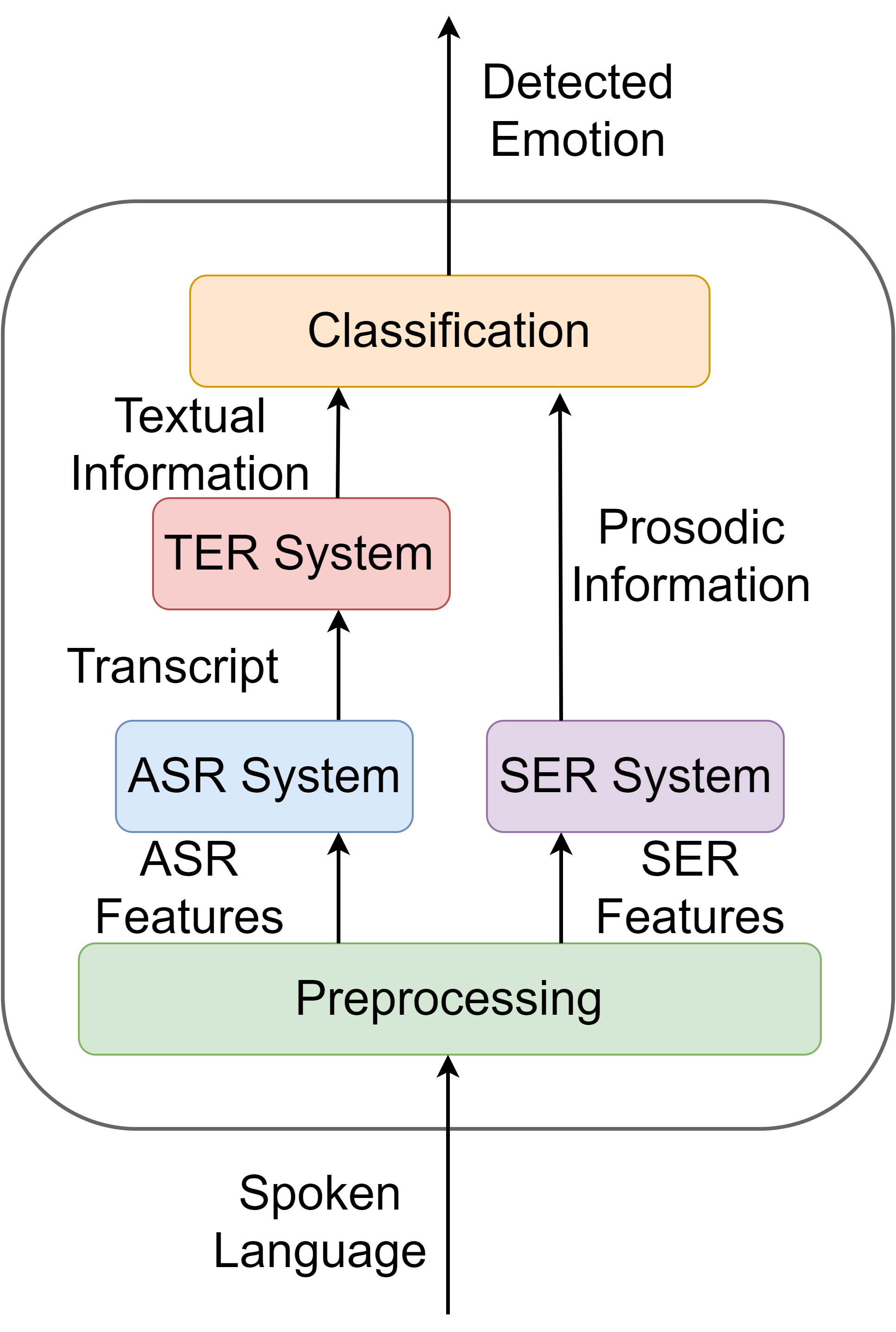}}
\end{minipage}
\caption{ The \emo \hspace{0.01cm} framework where each intermediate system (ASR, SER, TER) can be chosen and optimized individually.}
    \label{fig:embeddings}
\end{figure}

The \emo \hspace{0.01cm} framework consists of five parts. A given utterance is first forwarded into a pre-processing block, where both ASR and SER features are extracted from the same raw signal.
These features are then fed into an ASR and a (neural) SER System respectively. The ASR System automatically generates a transcript of the utterance, which is then used as an input for the (also neural) TER System. Both the spoken and the TER system then provide their outputs as independent features for the classification block. Finally, the classification block outputs the detected emotion. An overview can be seen in Figure \ref{fig:embeddings}.\par

This framework is very versatile and easily adaptable. The intermediate systems, i.e. the ASR, SER and TER systems, can be chosen and optimized independently, the only fixed variables being the input and the output size. Training \emo \hspace{0.01cm} is done in multiple steps. 
First the preprocessing block extracts ASR and SER features. Depending on the used sub-systems, the ASR features can range from classic MFCCs+Deltas to augmented waveforms as used in \cite{Park2019specaugment}. For the SER features either classical features such as pitch and prosody \cite{dellaert1996recognizing,Polzin1998Emotions}, complex emotion recognition feature sets (for example the IS09 set \cite{Schuller2009IS}) or augmented waveforms \cite{Sarma2018emotionIF} are plausible. \par 
After this, the utterances need to be transcribed to provide textual data. We found that modern out-of-the-box ASR systems generally provide high-quality transcriptions without the need for additional fine-tuning, allowing us to simply use one of these systems to transcribe the given utterances. Note that fine-tuning an ASR system towards a given dataset would probably provide slightly better transcriptions, leading to a slightly better performance overall. However, this comes at the cost of significantly more training time.\par
The SER and TER systems are trained individually on their parts of the given task, i.e. speech to emotion for SER and (possibly previously transcribed/generated) text to emotion for TER.
For this, both models are extended by an auxiliary classification block followed by a softmax layer, effectively creating two independent recognition systems.
\par
After this step is finished, we combine the SER and TER systems by removing the classification and softmax layers and replacing them with the classification block, leaving their second-to-last layers as the next block's input features.
The classification block usually consists of a simple feedforward network with a softmax classification layer. However, more sohpisticated approaches employing the attention mechanism might be better suited at aligning the two sets of features. The classification block is then trained while keeping the intermediate systems (ASR, SER and TER) unchanged. This is done to prevent potential overfitting due to huge relative differences in model size and structure.\par
A big advantage of this modular setup is that any intermediate system can simply be replaced by one another, requiring at worst the re-training of the classification block and the extraction of new/other features in the pre-processing block.
This proves to be especially useful considering the ongoing rapid development in both emotion recognition and deep learning in general. \emo \hspace {0.01cm} can adapt to these improvements by upgrading an outdated sub-system to a newer one without re-training all of the model.\par 

\section{Experimental Details / Results}
\label{sec:experiments}

\subsection{Pre-trained Language Models}
The performance of \emo \hspace{0.01cm} only relies on the ability of its intermediate systems to process sequence data. We found that using pre-trained language models for the TER part drastically improves performance while only taking very few training epochs to converge.
For our experiments, we use BERT \cite{Devlin2019bert} and XLNet \cite{Yang2019xlnet}, two popular transformer-based general language model that are available with pre-trained weights. They both consist of either 24 (Large) or 12 (Base) Encoder Transformer blocks and are pre-trained on large resources of textual data. For a detailed description of the Encoder Transformer block, see \cite{Vaswani2017attention}. We use the case-sensitive (cased) versions of both models.
Training BERT and XLNet on all tasks was conducted using the Adam-optimizer \cite{Kingma2014adam} with a learning rate of $3e-5$. 
We use the provided WordPiece embeddings for BERT and byte-pair encodings for XLNet. The sequence length is set to $48$. Longer/Shorter inputs are truncated/padded. We add a fully-connected layer of $768$ units with Dropout \cite{Srivastava2014Dropout} of $0.2$ and L2-regularization of $0.03$ between the pre-trained models and the softmax layer. We use weighted cross-entropy as our loss function to counteract the imbalanced datasets. These hyperparameters were chosen experimentally and work equally well on both SemEval and IEMOCAP.

\subsection{SemEval Experiments}
\label{subsec:SemEval}
We found that using pre-trained language models and fine-tuning them on emotional datasets achieves good results. In addition, fine-tuning instead of training from scratch significantly reduces training time and resource usage.
To show the effectiveness of pre-trained language models for TER, we run some tests on Task 4A of the SemEval 2017 competition \cite{Rosenthal2017semeval}, which is a 3-class sentiment analysis task.
The input is a tweet (a twitter post containing no more than 280 characters), the output is the predicted sentiment, in this case either `positive', `neutral', or `negative'. We choose this task because of its similarity to TER. The sentiments may be roughly mapped to the emotions `happy', `neutral' and  `sad'/`angry'.\par 
The dataset is hand-annotated and consists of about $50.000$ training and $12.000$ test samples consisting of tweets from 2012 to 2017. 
In the training set, the data consists of roughly $39.5/44.8/15.6\,\%$ positive/neutral/negative tweets while the test set has a distribution of $19.3/48.3/32.3\,\%$ for the respective classes.
In both cases, neutral tweets are the most common. However, the training data has a lot more positive tweets while the test set is more shifted towards negative ones.
This causes the task to be more challenging because the models need to detect negative sentiments with relatively little training data.\par 
\begin{table}
\begin{center}
\begin{tabular}{ c c } 
\textbf{Model} & \textbf{Accuracy} \\
\hline
All Neutral & $0.483$\\
\hline
DataStories \cite{Baziotis2017semeval} & $0.651$ \\
BB\_twtr \cite{Cliche2017semeval} & $0.658$ \\
LIA \cite{Rouvier2017lia} & $0.661$ \\ 
NNEMBs \cite{Yin2017nnembs} & $0.664$ \\
\hline
XLNet-Large & $0.663\pm0.030$\\
Bert-Base & $0.681\pm0.007$\\ 
Bert-Large & $0.687\pm0.015$\\ 
\textbf{XLNet-Base} & $\mathbf{0.695\pm0.007}$\\
\end{tabular}
\end{center}
\caption{Test accuracy on Task 4A of SemEval 2017. The best models are in bold. We repeat each experiment five times and report the mean and standard deviation.}
\label{tab:semeval}
\end{table}
 We experiment with both BERT and XLNet and find XLNet to be slightly superior in performance. The tweets were not pre-processed.
 \par 
 The results can be seen in Table \ref{tab:semeval}. We train each model five times with different initial weights and report the mean and standard deviation for each model. Our best model (XLNet-Base) achieves an accuracy of $69.5\,\%$ after only one epoch of training. This is an improvement of over $3\%$ absolute over previous models, showing the effectiveness of pre-trained models for this task.

\subsection{IEMOCAP Experiments}
The IEMOCAP dataset provides 12 hours of emotional audio dialogues, recorded with five pairs of professional actors in both scripted and spontaneous sessions.
At least three annotators were asked to label each of the dialogues using Ekman’s Base Emotions \cite{Ekman1999basic} and a variation of Russell’s Circumplex Model \cite{russell1980circumplex}.
The inter-annotator agreement for the task was 74.6\,\% (66.0\,\% for scripted and 83.1\,\% for spontaneous utterances) \cite{Busso2008iemocap}. Because of the scarcity of other emotions, we follow common research \cite{Yoon2018multimodal, Satt2017efficientER, Tripathi2018multimodal} and only classify the emotions `happiness', `sadness', `anger' and `neutral'. Our final dataset contains $38.04\,\%$ neutral, $24.57\,\%$ angry, $24.14\,\%$ sad and $13.24\,\%$ happy samples.
We split each dialogue into utterances from 3 to 15 seconds, with one utterance holding the information of one speaker (i.e. mono audio data and the respective transcription). Each utterance is then classified individually and without context information. We use both scripted and spontaneous dialogues for all experiments.\par 
We use 10-fold cross-validation with randomly shuffled data with a fixed random seed. Note that some researchers (e.g. \cite{Satt2017efficientER, Aguilar2019multimodal}) test their models in a speaker-exclusive setting instead.
They split the dataset by speakers, training on most speakers and testing exclusively on the remaining few. We denote the respective papers with a `*'.
We train each model three times with different initial weights and report the mean of all models and the mean of the three standard deviations.\par 
We evaluate the models with respect to weighted and unweighted accuracy. The Weighted Accuracy (WA) is given by the number of correctly classified samples divided by the total number of samples. 
The Unweighted Accuracy (UA) corresponds to first calculating the accuracy for each emotion class and then averaging by the number of samples per class.\par 
\subsubsection{IEMOCAP Text Experiments}
\label{sssec:iemocapText}
\begin{table}
\begin{center}
\begin{tabular}{ c c c c } 
\textbf{Model} & \textbf{WA} & \textbf{UA} & \textbf{T} \\
\hline 
TRE \cite{Yoon2018multimodal} & $0.635\pm0.018$ & -- & P \\
\textit{LSTM}* \cite{Tripathi2018multimodal} &$\mathit{0.648}$ & --& P\\
\hline
XLNet &  $0.694\pm0.015$ & $0.671\pm0.019$ & P \\
\textbf{BERT} & $\mathbf{0.709\pm0.015}$ & $\mathbf{0.691\pm0.018}$ & P\\
\hline
\hline
TRE-ASR \cite{Yoon2018multimodal} & $0.593\pm0.022$ & -- & A\\
\hline 
XLNet-A & $0.668\pm0.020$ & $0.645\pm0.023$ & A\\
\textbf{BERT-A} & $\mathbf{0.685\pm0.017}$ & $\mathbf{0.665\pm0.023}$ & A\\
\end{tabular}
\end{center}
\caption{Weighted Accuracy (WA) and Unweighted Accuracy (UA) for TER on the IEMOCAP dataset. Models without cross-validation are in \textit{cursive}. Models with `*' use speaker-exclusive testing. The T row denotes the Transcription. P stands for Provided and A for Automatic.}
\label{tab:IEMOText}
\end{table}

As described in Section \ref{sec:model}, we train \emo's \hspace{0.01cm} modalities independently at first. Motivated by their performance on the SemEval task (see Section \ref{subsec:SemEval}), we use both BERT and XLNet as TER models. We use the Base versions (i.e. 12 Transformer Encoder Blocks) for all experiments because they are roughly three times smaller than the large ones, making them more feasible for real-time applications. To compare the theoretical and the practical performance, we train on both provided and artificial ASR transcriptions \cite{Sperber2017Robust, ConstantinNW2019} with $6\,\%$ WER.
The latter are used as an alternative to an ASR system and are configured to simulate typical transcription errors as described in \cite{Sperber2017Robust}.
This WER was chosen to match the one reported in \cite{Yoon2018multimodal}. We use artificial noise instead of an ASR system due to time constraints to integrate an ASR of high quality.\par 
For all text-related experiments, we omitted the use of a validation split, instead training on a fixed number of seven epochs for all folds. The number seven was chosen because the training loss usually stabilized after seven epochs. The results can be seen in Table \ref{tab:IEMOText}. Here, BERT slightly outperforms XLNet, achieving an accuracy of $70.9\,\%$ on the provided transcripts and $68.5\,\%$ on the noisy ones.
While our models' performances still drops significantly when using noisy instead of provided transcriptions, they seem to be more robust to this noise than the one reported in \cite{Yoon2018multimodal}. We hypothesize that the extensive pre-training on large amounts of textual data leads to this increase in stability.\par

\subsubsection{IEMOCAP Speech Experiments}
\label{sssec:iemocapSpeech}
For the SER part, we train our models on the IS09 feature set \cite{Schuller2009IS}, following previous research \cite{Tarantino2019transformer, Ramet2018Attention}. The features were extracted using the openSMILE feature extraction tool \cite{Eyben2013OpenSmile} with a frame size of $300$ frames and a stride of $0.1$, resulting in an overlap of $0.9$ between two frames. We then pad/truncate the sequences to a fixed length of 384 frames.

The results in Table \ref{tab:IEMOSer} are achieved using simple Attention-BiLSTM models.
The models use a TDNN layer followed by two BiLSTM layers and an attention layer. The TDNN layer is used to map the inputs to the dimension of the BiLSTMs.
The output of the second BiLSTM layer is then fed into the attention layer. Then, the outputs of the second BiLSTM layer and the attention layer are concatenated and used as the input for a softmax layer.
Our best model, BiLSTM-128attDim, used a TDNN with filter size $16$, $64$ dimensional LSTMs (in both directions) and self-attention with dimension $128$. It achieved an accuracy of $60.8\,\%$. The second best model, BiLSTM-64attDim, uses the same configuration, except for an attention dimension of $64$, achieving $60.0\,\%$ weighted accuracy.
Note that more sophisticated SER models have been shown to yield far better performance and would thus be better suited for a combined bimodal architecture. However, we primarily want to showcase the capability of pre-trained language models for bimodal Emotion Recognition, making these basic models sufficient for this purpose.
\begin{table}
\begin{center}
\begin{tabular}{ c c c } 
\textbf{Model} & \textbf{WA} &\textbf{UA} \\
\hline
IS09-Self-Att \cite{Tarantino2019transformer} & $0.681$ & $\mathbf{0.638}$\\
\textbf{CNN-LSTM}*  \cite{Satt2017efficientER} & $\mathbf{0.688}$ & $0.594$\\
\textit{Att-TDNN/LSTM}* \cite{Sarma2018emotionIF} & $\mathit{0.701}$
 &$\mathit{0.607}$\\
\hline 
BiLSTM-64attDim & $0.600\pm0.020$  & $0.531\pm0.020$\\ 
BiLSTM-128attdim & $0.608\pm0.017$ & $0.526\pm0.013$ \\
\end{tabular}
\end{center}
\caption{Weighted Accuracy (WA) and Unweighted Accuracy (UA) on SER on the IEMOCAP dataset. Models without cross-validation are in \textit{cursive}. Models with `*' use speaker-exclusive testing.}
\label{tab:IEMOSer}
\end{table}
 \addtocounter{footnote}{-1}
\par 
\subsubsection{IEMOCAP Bimodal Experiments}
\label{sssec:iemocapBimodal}
\begin{table}
\begin{center}
\begin{tabular}{ c c c c } 
\textbf{Model} & \textbf{WA} & \textbf{UA} & \textbf{T} \\
\hline
H-MM-4* \cite{Aguilar2019multimodal} & $0.717$ & -- & P \\  
MDRE \cite{Yoon2018multimodal} & $0.718\pm0.019$& --&P \\
\textit{Att-align}* \cite{Xu2019multi} & $\mathit{0.725}$ & $\mathit{0.709}$ & P\\
\hline
\emo(XL) & $0.715\pm0.015$ & $0.689\pm0.016$ & P\\  
\textbf{\emo(BE)} & $\mathbf{0.735\pm0.016}$ & $\mathbf{0.710\pm0.016}$ & P\\
\hline
\hline
MDRE-ASR\cite{Yoon2018multimodal} & $0.691\pm0.019$ & -- &  A\\
\textit{Att-align-A}* \cite{Xu2019multi} & $\mathit{0.704}$ & $\mathit{0.695}$ & A\\
\hline
\emo(XL)-A &  $0.687\pm0.020$ & $0.657\pm0.020$ & A\\
\textbf{\emo(BE)-A} & $\mathbf{0.714\pm0.017}$ & $\mathbf{0.686\pm0.022}$ & A\\

\end{tabular}
\end{center}
\caption{Weighted Accuracy (WA) and Unweighted Accuracy (UA) on IEMOCAP using both speech and text data. Models without cross-validation are in \textit{cursive}. Models with `*' use speaker-exclusive testing. The T row denotes the Transcription. P stands for Provided and A for Automatic.}
\label{tab:IEMOfull}    
\end{table}
\par 
We combine our best individual models as described in Section \ref{sec:model}. Again, each model is trained for a fixed number of seven epochs. The results can be found in Table \ref{tab:IEMOfull}.
\emo(XL) \hspace{0.01cm} uses the best XLNet models from Table \ref{tab:IEMOText} as its text sub-system while \emo(BE) \hspace{0.01cm} uses the BERT models instead. Both combine them with the best SER model from Table \ref{tab:IEMOSer}.
We also tried fine-tuning the whole model (i.e. the classification block, the TER and the SER parts together) after the seven epochs.
However, the performance was significantly worse than the models that only trained the classification block. This is likely caused by the size of BERT and XLNet allowing them to easily dominate the classification outcome.  
The best models achieve $73.5\,\%$ and $71.4\,\%$
accuracy for provided and automatic transcripts respectively.
\par 

\section{Conclusion / Future Work}
\label{sec:conclusion}
Detecting emotions in speech is a challenging task that marks an important step towards more natural and adaptive human-computer interaction. We show that self-attentive pre-trained language models are well suited for text emotion recognition and that the Transformer model in general works well for bimodal Emotion Recognition.
Our experiments result in $69.5\,\%$ accuracy on Task 4A of SemEval 2017. We achieve $73.5\,\%$ and on the `Text+Speech' subtask of the IEMOCAP dataset. Using speech and (simulated) automatic transcriptions results in $71.4\,\%$ instead.
The results on IEMOCAP are achieved by fine-tuning BERT and combining it with a traditional speech emotion recognition system. This is done using \emo \hspace{0.01cm}, a modular framework that classifies both the speech data and a textual transcription in independent sub-modules that can be switched out without pre-training the whole model.\par
We suggest using Speech Emotion Recognition as a reward for Reinforcement Learning. It could be used to improve the quality of dialogue systems by allowing them to adjust their behaviour based on the emotional reactions of their dialogue partner.
We mention possible problems that arise due to different aspects of human emotion and propose a suitable reward function to circumvent them.\par 
Future work will involve the implementation of our framework into a Reinforcement Learning agent as well as more sophisticated combinations of Speech Emotion Recognition sub-modules. We will also repeat our experiments with an ASR system instead of artificial noise.
Another possible direction is to optimize pre-trained language models for emotional context, for example by adding an additional training step before fine-tuning them on a given task.

\section{Acknowledgements}
The project on which this report is based was funded by the Federal Ministry of Education and Research (BMBF) of Germany under the number 01IS18040A. The authors are responsible for the content of this publication.

\bibliographystyle{IEEEbib}
\bibliography{strings,refs}

\end{document}